# Secure Encryption scheme with key exchange for Two server Architecture

D.Siva Santosh Kumar
M.Tech, CNIS (IT)
VNRVJIET

Mrs.Kalyani
Assistant Professor, IT
kalyani_d@vnrvjiet.in
VNR VJIET

*Abstract*— In Distributed environment authentication and key-exchange mechanisms plays a major role. Generally in authentication a client and server agree upon a cryptographic key which is mutually exchanged. Earlier passwords were stored in a single server where if an intruder gains access of that server by some malicious attacks then all the passwords stored in the database are to be compromised and therefore the e-commerce application security is endangered. In-order to improve the efficiency and performance of the authentication mechanisms we involve multiple servers to store the passwords and participate in key-exchange and authentication process so as to ensure that even a single server is compromised the whole system's security is not in threat and also generating a nonce values randomly for every session corresponds to a two level security mechanism where the scope of an attacker to pretend as the legitimate user and login into the system is completely restricted.

In this model the passwords are stored in their equivalent hash values and they are spitted into multiple servers. Making it even more difficult for the attacker to determine the password even by reverse engineering he cannot intercept the actual password form segments of hash value. Hence, we can present this model as safe and secure password based authentication scheme using key exchange mechanism.

## I. INTRODUCTION

Passwords are basic entities of authentication and security, mainly in Distributed environment they play crucial role in authenticating legitimate users. The authentication systems which use passwords to authenticate are cost effective and are also easy to use. Users have to register with a password and remember it to login, irrespective of the device used. Passwords are strings or characters used to approve the access to gain privilege for a resource which is to be maintained confidential from other individuals who aren't given the access of a particular resource. The urge to develop a strong authentication system led to the advent of cryptography and many more authentication schemes like hierarchical identity based encryption scheme [6] with less no. of key exchanges. Authentication refers to sharing common information over a channel where they are compared and upon verified with those stored in database. If they match user is authenticated or-else reverted back prompting to provide with correct password in the other case.

Earlier passwords were hashed and their hash value was transmitted over a public channel leaving scope to an attacker to crack the password from hashed value by testing possible passwords and some advanced tools. Gradually, mechanisms are being developed which are capable of protecting passwords from eavesdropping or other guessing attacks by any means. In distributed computing environment these advanced techniques lead to the scenario where a client and a server can mutually authenticate by means of a password and a cryptographic key shared among them and this seem to be a primary and mandate feature where using key exchange mechanisms, server store the passwords (in terms of their equivalent hash values) and provide secure authentication for genuine and legitimate users prompting to login.

Key exchange refers to the mutual sharing of common password or credentials by establishing a cryptographic key to ensure any unauthorized user isn't allowed to participate in the session so as to prevent any brute force attacks on the credentials. User authentication through a single server has a risk that if the database is hacked by any means whole set of encrypted passwords are gained access and has a chance that they are used for illegitimate purposes thus leaving the server weak for using in password based authentication systems. The attacker by eaves-dropping the hash values communicated through any insecure channel can determine the password by trying with a wide number of possible passwords by brute-forcing and thus if a server is attacked and gained access it is possible to evaluate all the stored passwords and they are subjected to misuse.

The passwords stored in the server's database may also be sometimes exposed as they are easily prone to attacks where incase if the same password is used for other online service such as banking etc.., the confidentiality of credentials is being Compromised and due to a weak server the user's other online services are subjected to risk. This leads to continuous study and development of a secure authentication mechanism which further results in regular modifications as per the security requirements in distributed systems environment.

## II. RELATED WORK

In 1976 a cryptographic protocol was developed by Hellman's student Ralph Merkie which is published as a cryptographic protocol by Whitfield Diffie and Martin Hellman[1]. It enables its users to securely exchange keys even in-case of eavesdropping by third person. This exchange protocol however itself doesn't respond in authentication. It's crucial when a third person can eavesdrop at other end.

In-case of a single server if access to this server is gained through malicious attacks all the users passwords stored in the server's database are disclosed. To resolve this issue a PKI based model was proposed in 2000 by Ford and Kaliski where 'n' servers are involved in authenticating a client. It remained secure until only 'n-1' servers are attacked. Further a password only model was suggested with similar mechanism in 2001 by Joblon[5]. In 2002 another protocol in PKI model which requires only 'x' among 'n' servers to participate in authentication until 'x-1' servers remained secure was proposed by MacKenzie which provided a formal security for PAKE protocol in random oracle model[3]. A password only model was proposed in 2003 by Di Raimondo and Gennaro in which < 1/3 servers are required to be compromised with formal security model.

The first two-server model in PKI based setting was developed in 2003 by Brainard, in this a secure channel is assumed between the client and server which used public key technique implementations. The first two server password only authenticated key exchange protocol with a security standard model was proposed by Katz in 2005. This further extended and developed as Katz-Ostrovsky-Yung (KOY protocol)[2] and distributed key issuing protocol [7] based user identity which is taken as public key. In this protocol each server and client agree upon a common secret session key which is symmetric and two servers equally contribute in client authentication and their key exchange. Client action remains same whereas a server functions twice to thrice of its regular implementations against malicious attacks. Here two servers run in parallel which can be counted as an advantage despite its inefficiency in practical implementation. In-order to safeguard from a single point of threat a two server model was proposed in 2006. This model involved three elements users, Control server and a service server.

In 2007 a two server PAKE protocol with low number of round-off's in their communication was proposed by Jin. Here, client authenticates each other by verifying if they agree on common session key. In this paper we deal with a two server authentication PAKE protocol which makes both the servers to run in parallel with improved efficiency. We implement a two server scheme based on PAKE protocol where passwords are replaced by hash values by their equivalent hash values in-order to avoid plain text attack and also maintain confidentiality and they are stored in databases of both the servers by splitting[3]. Such that if a single server among 'n' servers is compromised too then also our system remains safe and secure leaving no scope for any malicious activities that could gain access of our system. The protocol is robust among offline dictionary attacks and also as an active adversary provided the parameters in need of control server to authenticate.

### III. CONTRIBUTION

This model deals with two-server authentication scheme using key exchange where the password among the multiple servers from the client side is stored in the form of their equivalent hash values and is split among the servers in an appropriate order. The intruder incase caught hold of one server may not be able to attack all the servers participating in authentication process and by any means if he does so, he may not intercept the exact password by reverse engineering the hash functions and re-arranging all the split values.

The scheme is initiated by firstly registration of user. It involves designating the username and password of the user to the server where the equivalent hash values are determined and they are spitted into 'n' segments and stored in 'n' servers for future reference. The number of spits is equal to number of servers available and it depends on the level of security the administrator wants to implement for his authentication system. When the user tries to login in with those credentials after successful submission the servers crosscheck with the respective usernames and the existing password corresponding to that username thus let us assume if two servers are participating in authentication then both the servers combine the hash-values and verify if the entered password hash value matches with that of the passwords stored in the servers. Thus upon successful verification the user is sent a nonce value randomly generated for that particular session. After this two step verification the server grants the access to the user declaring him as the legitimate user with respective privileges. User authentication implementation scheme result is as shown in figure1.

```
>>>
Welcome to the system. Kindly register to login.
Options: register/login/logout
> register
New username: Alex
New password: 0504
Creating account...
Account has been created
> register
New username: Rony
New password: 6451
Creating account...
Account has been created
> login
Username: Alex
Password: 0504
Login successful
Welcome to your account Alex
Options: logout
Alex > logout
Logging out...
> login
Username: Alex
Password: 6451
Invalid username or password
> |
```

The main aspect of this model is to ensure the system is safe even if one or more serves are compromised. As it is a tedious task to determine the password among the multiple servers by cracking the hash values. Even if the attacker tries with the replay attack the server can easily identify as he may not forge the nonce value which is randomly generated for every active session by the server. Still the process of developing a more susceptible model is being carried out in-order to improve the overall efficiency and performance resulting in a strong key-exchange mechanism involving multiple servers in distributed environment.

As per the analysis our model is safe and secure to most frequent prone attacks a more deeper analysis is to be done to ensure the system is completely safe and secure to withstand among the real-time existing protocols for authentication and key-exchange distributed environment.

### IV. SECURITY ANALYSIS

Our protocol ensures mutual authentication among the user and multiple servers and is presumed to be immune as far as tested with replay attack, offline dictionary attack, the authentication protocols proposed are prone to one or more security attacks once the server is compromised incase of single server systems but in our model as we involve multiple servers where the passwords in equivalent hash values are split and stored in multiple servers thereby reducing the scope of the server and our authentication scheme to be attacked or cracked by any means. As per the analysis we've done so far our model is resistant to some of the security breaches listed below

1. Impersonation attack:

In this attack, the attacker randomly assumes the security parameters and tries to access the system as a legitimate user. But In our protocol as we store the hash values of the passwords and they are split into servers it's a tedious task for an attacker to guess all the security parameters including the hash valued password and combine all its split values in respective order

2. Replay attack:

Here, the attacker intercepts all the logs exchanged between the user and servers and tried to replay the same communication pretending to be a legitimate user. But replaying the same messages is proven to be vague as we propose randomly generated sessions keys for each session in the authentication scheme.

3. Leak of Verifier attack:

The attacker by any means gain access to a server and eavesdrop all the messages communicated between the user and server but as we involve the participation of multiple servers in our model attacking a single server and gaining access to security parameters stored in that server isn't sufficient to be authenticated as a legitimate user.

4. Offline dictionary attack:

If the attacker eavesdrop the communication channel and tries to intercept the password by means of guessing and some password cracking tools he ends up in nothing as its very difficult to determine the password by reverse engineering a hash function exactly and determine the password by combining the password which is split into multiple servers

## CONCLUSION

This model serves the authentication and key-exchange mechanisms under a secure channel where incase if one or multiple (n-1) servers are compromised even then the intruder cannot determine the password as they are hashed and the hash values are split into multiple servers making it further more difficult for the intruder to crack the password. Thus, our mechanism justifies the concept of secure authentication and key exchange and as per the analysis the current model can withstand passive and active attacks and there is always scope for continuously improving and developing more advanced and strong defending mechanisms in authentication and key-exchange protocols.